\documentclass[aps,prd,nofootinbib,amsmath,amssymb,superscriptaddress,twocolumn,10pt]{revtex4}%superscriptaddress,showpacs,

\pdfoutput=1

\usepackage{graphicx}
\usepackage{dcolumn}
\usepackage{bm}
\usepackage{amssymb}
\usepackage{latexsym}
\usepackage{booktabs}
\usepackage{amsmath}
\usepackage{multirow}
\usepackage[colorlinks=true, linkcolor=red, citecolor=blue]{hyperref}

\newcommand{\be}{\begin{equation}}
\newcommand{\ee}{\end{equation}}
\newcommand{\bq}{\begin{eqnarray}}
\newcommand{\eq}{\end{eqnarray}}

\usepackage[usenames,dvipsnames]{xcolor}

\DeclareMathAlphabet\mathbfcal{OMS}{cmsy}{b}{n}
\definecolor{darkgreen}{cmyk}{0.85,0.2,1.00,0.2}
\definecolor{purple}{cmyk}{0.5,1.0,0,0}

\def\barray{\begin{array}}
\def\earray{\end{array}}
\def\be{\begin{equation}}
\def\ee{\end{equation}}
\def\ben{\begin{equation} \nonumber}
\def\een{\end{equation}}
\def\ban{\begin{eqnarray*}}
\def\ean{\end{eqnarray*}}
\def\ba{\begin{eqnarray}}
\def\ea{\end{eqnarray}}

\def\({\left(}
\def\){\right)}

\begin{document}

%\title{Weighing neutrinos in light of cosmological observations: Impacts from the simulated data of gravitational waves of the Einstein Telescope}

\title{Impacts of gravitational-wave standard siren observation of the Einstein Telescope on weighing neutrinos in cosmology}

\author{Ling-Feng Wang}
%\email{wanglf@stumail.neu.edu.cn}
\affiliation{Department of Physics, College of Sciences, Northeastern University, Shenyang
110004, China}
\author{Xuan-Neng Zhang}
%\email{1677019892@qq.com}
\affiliation{Department of Physics, College of Sciences, Northeastern University, Shenyang
110004, China}
\author{Jing-Fei Zhang}
%\email{jfzhang@mail.neu.edu.cn}
\affiliation{Department of Physics, College of Sciences, Northeastern University, Shenyang
110004, China}
\author{Xin Zhang\footnote{Corresponding author}}
\email{zhangxin@mail.neu.edu.cn}
\affiliation{Department of Physics, College of Sciences, Northeastern University, Shenyang 110004, China}
\affiliation{Center for High Energy Physics, Peking University, Beijing 100080, China}

\begin{abstract}

We investigate the impacts of the gravitational-wave (GW) standard siren observation of the Einstein Telescope (ET) on constraining the total neutrino mass. We simulate 1000 GW events that would be observed by the ET in its 10-year observation by taking the standard $\Lambda$CDM cosmology as a fiducial model. We combine the simulated GW data with other cosmological observations including cosmic microwave background (CMB), baryon acoustic oscillations (BAO), and type Ia supernovae (SN). We consider three mass hierarchy cases for the neutrino mass, i.e., normal hierarchy (NH), inverted hierarchy (IH), and degenerate hierarchy (DH). Using Planck+BAO+SN, we obtain $\sum m_\nu<0.175$ eV for the NH case, $\sum m_\nu<0.200$ eV for the IH case, and $\sum m_\nu<0.136$ eV for the DH case. After considering the GW data, i.e., using Planck+BAO+SN+GW, the constraint results become $\sum m_\nu<0.151$ eV for the NH case, $\sum m_\nu<0.185$ eV for the IH case, and $\sum m_\nu<0.122$ eV for the DH case. We find that the GW data can help reduce the upper limits of $\sum m_\nu$ by 13.7\%, 7.5\%, and 10.3\% for the NH, IH, and DH cases, respectively. In addition, we find that the GW data can also help break the degeneracies between $\sum m_{\nu}$ and other parameters. We show that the GW data of the ET could greatly improve the constraint accuracies of cosmological parameters.

\end{abstract}
%\pacs{95.36.+x, 98.80.Es, 98.80.-k}
\maketitle

\section{Introduction}
\label{sec1}
On 17 August 2017, the signal of a gravitational wave (GW) produced by the merger of a binary neutron-star system (BNS) was detected for the first time \cite{TheLIGOScientific:2017qsa}, and the electromagnetic (EM) signals generated by the same transient source were also observed subsequently, which indicates that the age of gravitational-wave multi-messenger astronomy is coming. The measurement of GWs from the merger of a binary compact-object system involves the information of absolute luminosity distance of the transient source \cite{Schutz:1986gp}, and thus if we can simultaneously accurately measure the GW and EM signals from the same merger event of a BNS or a binary system consisting of a neutron star (NS) and a black hole (BH), then we are able to establish a true luminosity distance--redshift ($d_{L}$--$z$) relation. Therefore, the GW observations can serve as a cosmic ``standard siren'', which can be developed to be a new cosmological probe if we can accurately observe a large number of merger events of this class.

The current mainstream cosmological probes include the measurements of cosmic microwave background (CMB) anisotropies (temperature and polarization), baryon acoustic oscillations (BAO), type Ia supernovae (SNIa), and the Hubble constant, etc. In addition, there are also some observations for the growth history of large-scale structure (LSS), such as the shear measurement of the weak gravitational lensing, the galaxy clusters number counts in light of Sunyaev-Zeldovich (SZ) effect, and the CMB lensing measurement, etc. When using these observational data to make cosmological parameter estimation, some problems occur including mainly: (i) there are degeneracies between some parameters, and in some of which the correlations are rather strong, and (ii) there are apparent tensions between some observations. The GW standard siren observations have some peculiar advantages in breaking the parameter degeneracies, owing to the fact that the GW observations can directly measure the true luminosity distances, but the SNIa observations actually can only measure the ratios of luminosity distances at different redshifts, but not the true luminosity distances. In addition, compared to the standard candle provided by the SNIa observations \cite{Riess:1998cb,Perlmutter:1998np,Suzuki:2011hu}, which needs to cross-calibrate the distance indicators on different scales, the GW observations allow us to directly measure the luminosity distances up to higher redshifts \cite{Tamanini:2016zlh}. To obtain the information of redshifts, one needs to detect the EM counterparts of the GW sources. In fact, there are some forthcoming large facility observation programs, such as Large Synoptic Survey Telescope (LSST) \cite{LSST}, Square Kilometer Array (SKA) \cite{SKA}, and Extremely Large Telescope (ELT) \cite{ELT}, which can detect the EM counterparts by optically identifying the host galaxies. In this way, in the near future, the $d_{L}$--$z$ relation would be obtained by the GW standard siren measurements, and we could use this powerful tool to explore the expansion history of the universe.

Recently, the related issues have been discussed by some authors \cite{Cai:2016sby,Cai:2017aea,Cai:2017buj,Cai:2017yww,Sathyaprakash:2009xt,Zhao:2010sz,Li:2013lza,Yang:2017bkv,Feeney:2018mkj,Liao:2017ioi} (see also Ref.~\cite{Cai:2017cbj} for a recent review). For example, in Ref.~\cite{Cai:2016sby}, Cai and Yang estimated the ability of using GW data to constrain cosmological parameters. They considered to use the GW detector under planning, the Einstein Telescope (ET), to simulate the GW data, which is a third-generation ground-based detector of GWs \cite{ET}. ET is ten times more sensitive than the current advanced ground-based detectors and it covers the frequency range of $1-10^{4}$ Hz. According to their results \cite{Cai:2016sby}, the errors of cosmological parameters can be constrained to be $\Delta h\sim 5\times 10^{-3}$ and $\Delta \Omega_m\sim 0.02$ when using 1000 GW events, whose sensitivity is comparable to that of the Planck data \cite{Ade:2015xua}. From their study, we can see that the GW data can indeed be used to improve the constraint accuracies of parameters. In Ref.~\cite{etnew}, we can see how the parameter degeneracies are broken by the GW observations in an efficient way. In this work, we investigate the issue of measuring the neutrino mass in light of cosmological observations and we will discuss what role the GW observations of the ET will play as a new cosmological probe in this study.

Since the phenomenon of neutrino oscillation was discovered, which proved that the neutrino masses are not zero \cite{Lesgourgues:2006nd}, the determination of neutrino masses has been an important issue in the field of particle physics. Due to the fact that neutrino oscillation experiments are only sensitive to the squared mass differences between the neutrino mass eigenstates, it is a great challenge to determine the absolute masses of neutrinos by particle physics experiments. Although the solar and atmospheric neutrino oscillation experiments can give two squared mass differences between the mass eigenstates: $\Delta m_{21}^2\simeq 7.5\times 10^{-5}$ eV$^2$ and $|\Delta m_{32}^2|\simeq 2.5 \times 10^{-3}$ eV$^2$, we cannot determine whether the third neutrino is heaviest or lightest. Therefore, these measurements can only give two possible mass orders, i.e., the normal hierarchy (NH) with $m_1<m_2\ll m_3$ and the inverted hierarchy (IH) with $m_3\ll m_1<m_2$. If the total mass of neutrinos ($\sum m_{\nu}$) can be measured, then the absolute masses of neutrinos could be solved by combining the total mass and the two squared mass differences.

Massive neutrinos play an significant role in the evolution of the universe, and thus they leave distinct signatures on CMB and LSS at different epochs of the evolution of the universe. These signatures can actually be extracted from the cosmological observations, from which the total mass of neutrinos can be effectively constrained. In recent years, the combinations of various cosmological observations have been providing more and more tight constraint limits for the total mass of neutrinos. For the latest progresses on this aspect, see e.g. Refs.~\cite{Zhang:2017rbg,Guo:2017hea,Zhao:2016ecj,Zhang:2015uhk,Huang:2015wrx,Wang:2016tsz,Zhao:2017jma,Feng:2017mfs,Zhao:2017urm,Feng:2017nss,Zhang:2015rha,Zhang:2014ifa,Vagnozzi:2018jhn,Lorenz:2017fgo,Vagnozzi:2017ovm,Capozzi:2017ipn,Capozzi:2016rtj,Yang:2017amu,Xu:2016ddc}.

In fact, in a recent paper \cite{DiValentino:2017clw}, the constraints on the total neutrino mass have been discussed by considering the inclusion of the actual observation of GW and EM emission produced by the merger of BNS, GW170817. In Ref.~\cite{DiValentino:2017clw}, the authors considered a 12-parameter extended cosmological model that contains the dark-energy equation-of-state parameter $w_0$ and $w_a$, the spatial curvature $\Omega_k$, the total neutrino mass $\sum m_\nu$, the effective number of relativistic species $N_{\rm eff}$, and the running of the scalar spectral index $dn_s/d\ln k$, besides the 6 base parameters, and they made a comparison for the constraint results of Planck and Planck+GW170817. For such a 12-parameter model, using only Planck data gives $\sum m_\nu<1.11$ eV, and the combination of Planck+GW170817 gives $\sum m_\nu<0.77$ eV, showing that the inclusion of GW170817 leads to an about 30\% improvement for the upper limit of $\sum m_\nu$, compared to the result obtained from the Planck data alone. However, it is clearly known that the 12-parameter model actually cannot be well constrained by the current observations. In the present work, we wish to scrutinize the standard cosmology, i.e., a 7-parameter $\Lambda$ cold dark matter ($\Lambda$CDM) model that contains the 6 base parameters plus $\sum m_\nu$. We aim to see how the future GW observations help improve the constraints on the total neutrino mass in a 7-parameter $\Lambda$CDM cosmology.

%Recently, the constraint ability on the total neutrino mass of GW170817 has been discussed in \cite{DiValentino:2017clw}. In that work, the model considered is the CPL model where the dark energy component is parameterized using the Chevalier-Polarski-Linder parametrization. Compared with the case of only using Planck data, the ``GW170817" helps further constrain the total neutrino mass from $\sum m_\nu<1.11$ eV to $\sum m_\nu<0.77$ eV. The cosmological model considered in that work contains 12 parameters. Adding extra parameters will generally lead to a looser constraint on the total neutrino mass, so the constraint on the total neutrino mass will improve a lot in this case. However, we want to investigate the extent to which gravitational waves can improve the constraint on $\sum m_\nu$ in a model where the $\sum m_\nu$ can be well constrained. So we consider a 7-parameter model in our work, which contains 6 parameters in the standard $\Lambda$CDM model and parameter of total neutrino mass $\sum m_\nu$. Because the BAO and SN actually can help to break the degeneracy between $\sum m_\nu$ and other parameter, so we also add BAO and SN in our data sets. In addition, only a few gravitational wave events have been observed until now. While we want to know how more gravitational wave data will improve the constraint ability on cosmological parameters after a few years, this requires us to simulate future GW observations.

In this study, our main focus is on the ability of the GW observations of ET to constrain the neutrino mass. We follow Ref.~\cite{Cai:2016sby} to generate the catalogue of GW events, from which we can get the corresponding $d_{L}$--$z$ relation. According to the influences of instruments and weak lensing, we can estimate the uncertainty on the measurement of $d_{L}$. We simulate 1,000 GW events that could be observed by the ET in its 10-year observation. Then, we combine these simulated GW data with other current observations to constrain the total mass of neutrinos, $\sum m_{\nu}$, by using the Markov-chain Monte Carlo (MCMC) \cite{Lewis:2002ah} approach.

The paper is organized as follows. In Sec.~\ref{sec2}, we describe the method to generate simulated GW events and get the luminosity distances with the simulated measurement errors. In addition, the fiducial model and the data processing method are also briefly introduced. In Sec.~\ref{sec3}, we report the results of this work. In Sec.~\ref{sec4}, we make a conclusion for this work.

\section{Methods of simulating data and constraining parameters}
\label{sec2}

\subsection{Method of simulating data}
The first step for generating GW data is to simulate the redshift distribution of the sources. Following Refs.~\cite{Cai:2016sby,Zhao:2010sz}, the distribution takes the form
\begin{equation}
P(z)\propto \frac{4\pi d_C^2(z)R(z)}{H(z)(1+z)},
\label{equa:pz}
\end{equation}
where $d_C(z)$ is the comoving distance at redshift $z$ and $R(z)$ denotes the time evolution of the burst rate and takes the form \cite{Schneider:2000sg,Cutler:2009qv,Cai:2016sby}
\begin{equation}
R(z)=\begin{cases}
1+2z, & z\leq 1, \\
\frac{3}{4}(5-z), & 1<z<5, \\
0, & z\geq 5.
\end{cases}
\label{equa:rz}
\end{equation}
According to the prediction of the Advanced LIGO-Virgo network, we take the ratio between BHNS (the binary system of a BH and a NS) and BNS events to be 0.03. For the mass distributions of NS and BH in the simulation, we randomly sample the mass of NS in the interval $[1,2]~M_{\odot}$ and the mass of BH in the interval $[3,10]~M_{\odot}$ (here $M_{\odot}$ is the solar mass), as the same as in Ref.~\cite{Cai:2016sby}.

%We take the masses distribution to be in the interval [1,2] $M_{\odot}$ for NS and [3,10] $M_{\odot}$ for BH (here $M_{\odot}$ is the solar mass), as the same as in Ref.~\cite{Cai:2016sby}.

After the distribution of the sources is known, the second step is to generate the catalogue of the GW sources through the fiducial model, i.e., the $\Lambda$CDM model. If we consider a flat Friedmann-Robertson-Walker universe, the Hubble parameter $H(z)$ for the $\Lambda$CDM cosmology can be written as
\begin{align}
H(z)^2 &= H_0^2 \left[ (1 - {\Omega _m} - {\Omega _r}) + {\Omega _m}{(1 + z)^3} + \right. \nonumber\\
&\quad\left.{\Omega _r}{(1 + z)^4}\right],
\end{align}
where $\Omega _m$ and $\Omega_r$ represent the fractional energy densities of matter and radiation, respectively. In the late universe, $\Omega_r$ can be neglected, and $\Omega _m=\Omega_{b}+\Omega _c+\Omega _\nu$, where $\Omega_{b}$, $\Omega _c$, and $\Omega _\nu$ represent the fractional energy densities of baryons, cold dark matter, and neutrinos, respectively. Note that $\Omega _\nu$ can be expressed as
\begin{align}
\Omega _\nu = \frac{\sum m_{\nu}}{94h^{2}{\rm eV}},
\end{align}
where $h$ is the reduced Hubble constant (the Hubble constant $H_{0}= 100h$ $\rm km~s^{-1}~Mpc^{-1}$). Since the radiation component can be ignored in the late universe, we set $\Omega _r=0$ here.

The luminosity distance $d_L$ can be calculated by
\begin{equation}
{d_L} = \frac{{(1 + z)}}{{H_0}}\int_0^z {\frac{{dz'}}{{E(z')}}},
\label{equa:dl}
\end{equation}
where $E(z)\equiv H(z)/H_0$. According to the redshift distribution of the GW sources, we can then use Eq.~(\ref{equa:dl}) to generate a catalogue of the GW sources. That is to say, the $d_{L}$--$z$ relation can be obtained for every GW event in a $\Lambda$CDM cosmology.

The next step is to get the error of $d_{L}$ of the GW source, which is denoted by $\sigma_{d_{L}}$ in this paper. We first need to generate the waveform of GWs. Because the GW amplitude depends on the luminosity distance $d_{L}$, we can get the information of $d_{L}$ and $\sigma_{d_{L}}$ from the amplitude of waveform.

Following Ref.~\cite{Cai:2016sby}, the strain in GW interferometers can be written as
\begin{equation}
h(t)=F_+(\theta, \phi, \psi)h_+(t)+F_\times(\theta, \phi, \psi)h_\times(t),
\end{equation}
where $\psi$ is the polarization angle and ($\theta$,$\phi$) are angles describing the location of the source relative to the detector. Here the antenna pattern functions of the ET (i.e. $F_{+}$ and $F_{\times}$) are \cite{Zhao:2010sz}
\begin{align}
F_+^{(1)}(\theta, \phi, \psi)=&~~\frac{{\sqrt 3 }}{2}[\frac{1}{2}(1 + {\cos ^2}(\theta ))\cos (2\phi )\cos (2\psi ) \nonumber\\
                              &~~- \cos (\theta )\sin (2\phi )\sin (2\psi )],\nonumber\\
F_\times^{(1)}(\theta, \phi, \psi)=&~~\frac{{\sqrt 3 }}{2}[\frac{1}{2}(1 + {\cos ^2}(\theta ))\cos (2\phi )\sin (2\psi ) \nonumber\\
                              &~~+ \cos (\theta )\sin (2\phi )\cos (2\psi )].
\label{equa:F}
\end{align}
For the other two interferometers, their antenna pattern functions can be derived from above equations because the interferometers have $60^\circ$ with each other.

Following Refs.~\cite{Zhao:2010sz,Li:2013lza}, we can compute the Fourier transform $\mathcal{H}(f)$ of the time domain waveform $h(t)$,
\begin{align}
\mathcal{H}(f)=\mathcal{A}f^{-7/6}\exp[i(2\pi ft_0-\pi/4+2\psi(f/2)-\varphi_{(2.0)})],
\label{equa:hf}
\end{align}
where $\mathcal{A}$ is the Fourier amplitude that is given by
\begin{align}
\mathcal{A}=&~~\frac{1}{d_L}\sqrt{F_+^2(1+\cos^2(\iota))^2+4F_\times^2\cos^2(\iota)}\nonumber\\
            &~~\times \sqrt{5\pi/96}\pi^{-7/6}\mathcal{M}_c^{5/6},
\label{equa:A}
\end{align}
where $\mathcal{M}_c=M \eta^{3/5}$ is the ``chirp mass", $M=m_1+m_2$ is the total mass of coalescing binary with component masses $m_1$ and $m_2$, and $\eta=m_1 m_2/M^2$ is the symmetric mass ratio. Note that all the masses here refer to the observed mass rather than the intrinsic mass. The observed mass is larger than the intrinsic mass by a factor of $(1+z)$, i.e., $M_{\rm obs}=(1+z)M_{\rm int}$. $\iota$ is the angle of inclination of the binary's orbital angular momentum with the line of sight. Due to the fact that the short gamma ray bursts (SGRBs) are expected to be strongly beamed, the coincidence observations of SGRBs imply that the binaries should be orientated nearly face on (i.e., $\iota\simeq 0$) and the maximal inclination is about $\iota=20^\circ$. In fact, averaging the Fisher matrix over the inclination $\iota$ and the polarization $\psi$ with the constraint $\iota<90^\circ$ is roughly the same as taking $\iota=0$ in the simulation \cite{Li:2013lza}. Thus, when we simulate the GW source we can take $\iota=0$. But, when we estimate the practical uncertainty of the measurement of $d_L$, the impacts of the uncertainty of inclination should be taken into account. Actually, the consideration of the maximal effect of the inclination (between $\iota=0$ and $\iota=90^\circ$) on the signal-to-noise (SNR) leads to a factor of 2 [see Eq.~(\ref{sigmainst})]. The definitions of other parameters and the values of the parameters can be found in Ref.~\cite{Cai:2016sby}.

%When we simulate the GW source we consider the ideal situation where $\iota=0$, while the inclination is taken into account when we estimate the uncertainty of the measurement of   $d_L$. The definitions of other parameters and the values of the parameters can be found in Ref.~\cite{Cai:2016sby}.

%In line with the previous work of simulating future gravitational wave data, such as Refs.~\cite{Cai:2016sby,Zhao:2010sz,Cai:2017aea,Yang:2017bkv}, we select gravitational wave events simulated with SNR$>$8. The significance of this step is that a GW source is visible if it produces an SNR of at least 8 in ET \cite{ET}. 

After knowing the waveform of GWs, we can then calculate the SNR. A GW detection is confirmed if it produces a combined SNR of at least 8 in ET \cite{ET,Sathyaprakash:2012jk} (see also Refs.~\cite{Cai:2016sby,Zhao:2010sz,Cai:2017aea,Yang:2017bkv}). The combined SNR for the network of three independent interferometers is
\begin{equation}
\rho=\sqrt{\sum\limits_{i=1}^{3}(\rho^{(i)})^2},
\label{euqa:rho}
\end{equation}
where $\rho^{(i)}=\sqrt{\left\langle \mathcal{H}^{(i)},\mathcal{H}^{(i)}\right\rangle}$, with the inner product being defined as \begin{equation}
\left\langle{a,b}\right\rangle=4\int_{f_{\rm lower}}^{f_{\rm upper}}\frac{\tilde a(f)\tilde b^\ast(f)+\tilde a^\ast(f)\tilde b(f)}{2}\frac{df}{S_h(f)},
\label{euqa:product}
\end{equation}
where a ``$\sim$" above a function denotes the Fourier transform of the function and $S_h(f)$ is the one-side noise power spectral density. In this work, we take $S_h(f)$ of the ET to be the same as in Ref.~\cite{Zhao:2010sz}.

Using the Fisher information matrix, we can estimate the instrumental error on the measurement of $d_{L}$, which can be written as
\begin{align}
\sigma_{d_L}^{\rm inst}\simeq \sqrt{\left\langle\frac{\partial \mathcal H}{\partial d_L},\frac{\partial \mathcal H}{\partial d_L}\right\rangle^{-1}}.
\end{align}
Because we only focus on the parameter $d_{L}$ in the waveform, we find that $\sigma_{d_L}^{\rm inst}\simeq d_L/\rho$ due to $\mathcal H \propto d_L^{-1}$. Considering the effect from the inclination angle $\iota$, we add a factor 2 in front of the error, so the error is written as
\begin{equation}
\sigma_{d_L}^{\rm inst}\simeq \frac{2d_L}{\rho}.
\label{sigmainst}
\end{equation}
Following Ref.~\cite{Cai:2016sby}, we set the additional error $\sigma_{d_L}^{\rm lens}$ = $0.05z d_L$, which represents the error from weak lensing. Thus, the total error on the measurement of $d_{L}$ can be expressed as
\begin{align}
\sigma_{d_L}&~~=\sqrt{(\sigma_{d_L}^{\rm inst})^2+(\sigma_{d_L}^{\rm lens})^2} \nonumber\\
            &~~=\sqrt{\left(\frac{2d_L}{\rho}\right)^2+(0.05z d_L)^2}.
\label{sigmadl}
\end{align}

Using the method described above, we can generate the catalogue of the GW events with their $z$, $d_{L}$, and $\sigma_{d_{L}}$. According to the results in Ref.~\cite{Cai:2016sby}, we know that it requires more than 1000 GW events to match the Planck sensitivity, and so we simulate 1000 GW events that are expected to be detected by the ET in its 10-year observation.

\subsection{Method of constraining parameters}

In order to constrain cosmological parameters, we use the MCMC method to infer the posterior probability distributions of parameters. To measure the total neutrino mass in light of cosmological observations, we set $\sum m_{\nu}$ to be a free parameter in the cosmological fit.

In this work, we add the simulated GW data in the combined cosmological data. For the GW standard siren measurement with $N$ simulated data points, we can write its $\chi^2$ as
\begin{align}
\chi_{\rm GW}^2=\sum\limits_{i=1}^{N}\left[\frac{\bar{d}_L^i-d_L(\bar{z}_i;\vec{\Omega})}{\bar{\sigma}_{d_L}^i}\right]^2,
\label{equa:chi2}
\end{align}
where $\bar{z}_i$, $\bar{d}_L^i$, and $\bar{\sigma}_{d_L}^i$ are the $i$th redshift, luminosity distance, and error of luminosity distance of the simulated GW data, and $\vec{\Omega}$ represents the set of cosmological parameters.

To show the constraining capability of the simulated GW data, we consider two data combinations for comparison in this work: (i) Planck+BAO+SN and (ii) Planck+BAO+SN+GW. For the CMB data, we use the Planck 2015 temperature and polarization data. For the BAO data, we use the measurements of the six-degree-field galaxy (6dFGS) at $z_{\rm{eff}} = 0.106$ \cite{Beutler:2011hx}, the SDSS main galaxy sample (MGS) at $z_{\rm{eff}} = 0.15$ \cite{Ross:2014qpa}, the baryon oscillation spectroscopic survey (BOSS) LOWZ at $z_{\rm{eff}} = 0.32$ \cite{Anderson:2013zyy}, and the BOSS CMASS at $z_{\rm{eff}} = 0.57$ \cite{Anderson:2013zyy}. For the SN data, we use the ``joint light-curve analysis" (JLA) sample \cite{Betoule:2014frx}. For the simulated GW data, we consider 1,000 GW events that could be observed by the ET in its 10-year observation.

For the neutrino mass measurement in this work, we consider three mass hierarchy cases, i.e., the normal hierarchy (NH), the inverted hierarchy (IH), and the degenerate hierarchy (DH). For the details of this aspect, see Refs.~\cite{Wang:2016tsz,Huang:2015wrx}.

\section{Results and discussion}\label{sec3}

\begin{table*}[!htp]
\caption{Constraint results of the cosmological parameters by using the Planck+BAO+SN data (Data1) and the Planck+BAO+SN+GW data (Data2). Note that $\sum m_\nu$ is in units of eV and $H_0$ is in units of $\rm km~s^{-1}~ Mpc^{-1}$.}~\\
\centering
\label{tab1}
\renewcommand{\arraystretch}{1.5}
\scalebox{1}[1]{%
\begin{tabular}{|c|c c|c c|c c|}
\hline
&\multicolumn{2}{c|}{NH}  &  \multicolumn{2}{c|}{IH}  &  \multicolumn{2}{c|}{DH}     \\
\cline{2-7}
&Data$1$&Data$2$&Data$1$&Data$2$&Data$1$&Data$2$\\
\hline
$\Omega_bh^2$&$0.02235\pm0.00014$&$0.02234^{+0.00011}_{-0.00012}$&$0.02236\pm0.00014$&$0.02233^{+0.00011}_{-0.00012}$&$0.02233\pm0.00014$&$0.02229^{+0.00012}_{-0.00013}$\\
 $\Omega_ch^2$&$0.1183^{+0.0011}_{-0.0010}$&$0.1187^{+0.0008}_{-0.0006}$&$0.1180\pm0.0010$&$0.1187^{+0.0008}_{-0.0006}$&$0.1187\pm0.0011$&$0.1190^{+0.0010}_{-0.0007}$\\
 $100\theta_{\rm{MC}}$&$1.04093^{+0.00030}_{-0.00029}$&$1.04093\pm0.00027$&$1.04094\pm0.00030$&$1.04092^{+0.00026}_{-0.00027}$&$1.04091\pm0.00029$&$1.04086\pm0.00028$\\
 $\tau$&$0.089\pm0.017$&$0.086\pm0.016$&$0.091\pm0.016$&$0.088\pm0.016$&$0.084^{+0.017}_{-0.016}$&$0.082\pm0.016$\\
 ${\rm{ln}}(10^{10}A_s)$&$3.109^{+0.033}_{-0.032}$&$3.105\pm0.032$&$3.113\pm0.032$&$3.109\pm0.032$&$3.100^{+0.033}_{-0.032}$&$3.097\pm0.032$\\
 $n_s$&$0.9685^{+0.0041}_{-0.0042}$&$0.9675^{+0.0034}_{-0.0038}$&$0.9692\pm0.0040$&$0.9675\pm0.0035$&$0.9675\pm0.0041$&$0.9666\pm0.0036$\\
\hline
$\Omega_m$&$0.3098^{+0.0064}_{-0.0065}$&$0.3117\pm0.0024$&$0.3120\pm0.0064$&$0.3158\pm0.0026$&$0.3069^{+0.0065}_{-0.0071}$&$0.3090\pm0.0023$\\
 $H_0$&$67.64^{+0.50}_{-0.49}$&$67.52\pm0.16$&$67.43\pm0.49$&$67.18\pm0.17$&$67.92\pm0.52$&$67.75\pm0.14$\\
 \hline
 $\sum m_\nu$&$<0.175$&$<0.151$&$<0.200$&$<0.185$&$<0.136$&$<0.122$\\
\hline
\end{tabular}}

\end{table*}

\begin{figure*}[!htp]
\includegraphics[scale=0.65]{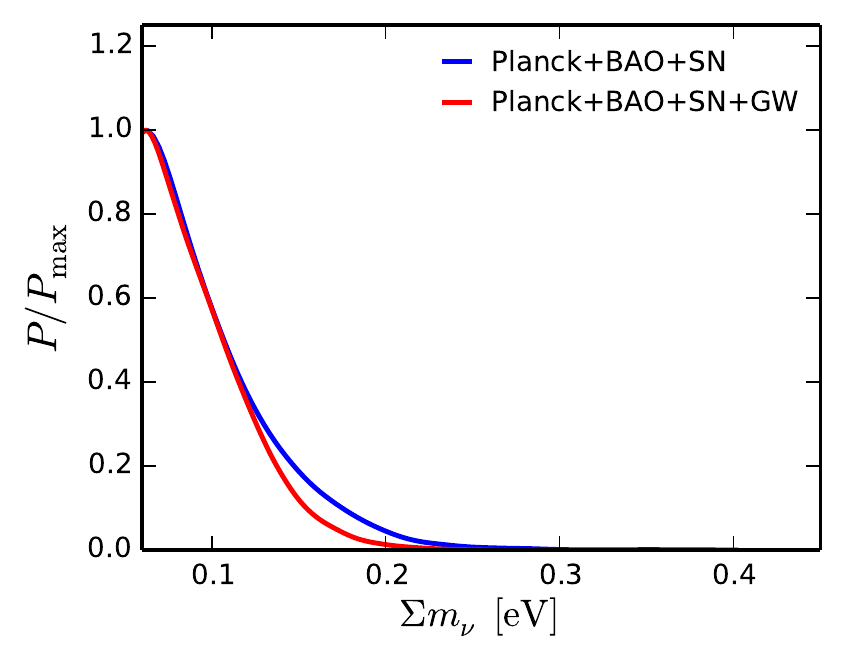}
\includegraphics[scale=0.65]{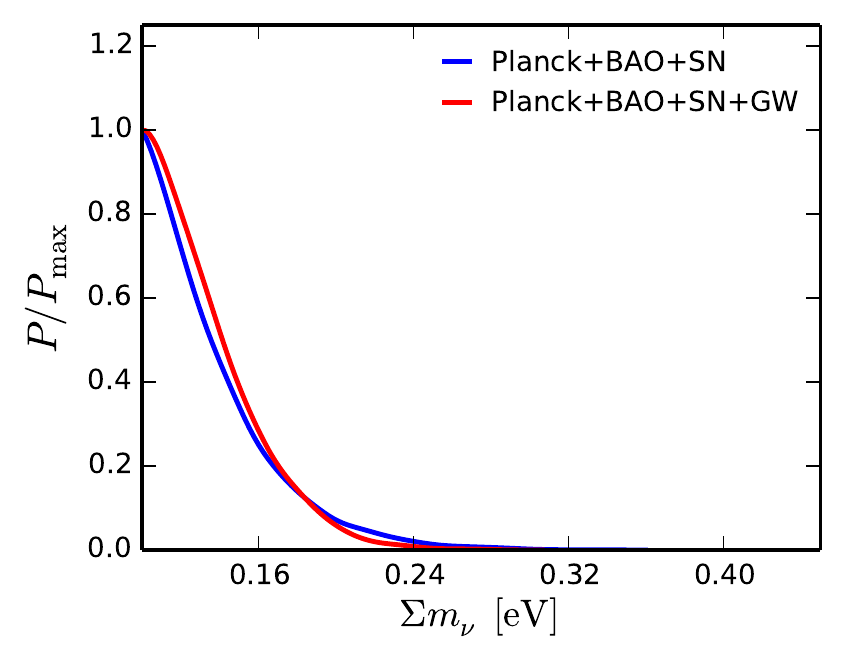}
\includegraphics[scale=0.65]{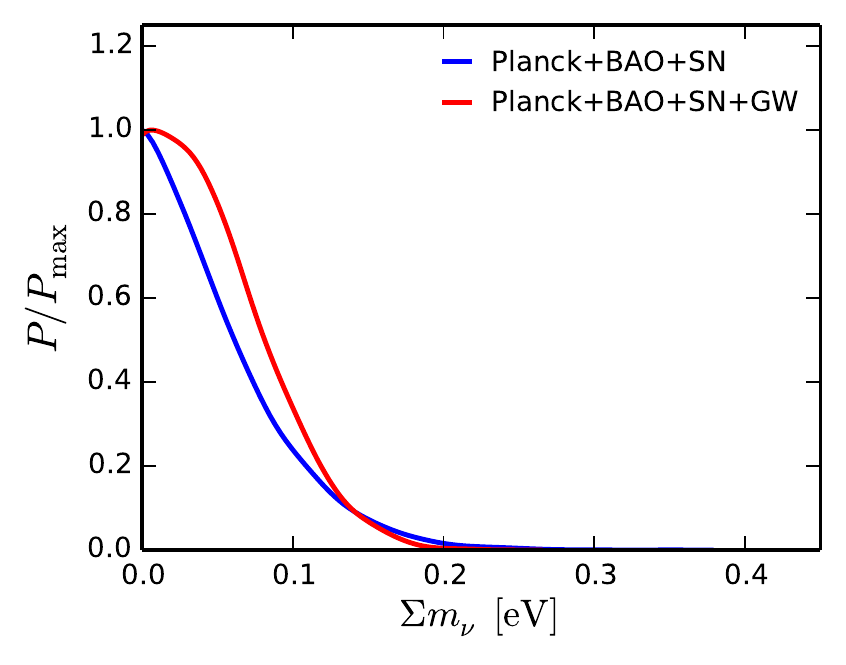}
\centering
 \caption{\label{fig1} The one-dimensional marginalized distributions of $\sum m_{\nu}$ using Planck+BAO+SN ({\it blue}) and Planck+BAO+SN+GW ({\it red}). ({\it left}) NH; ({\it middle}) IH; ({\it right}) DH.}
\end{figure*}

\begin{figure*}[!htp]
\includegraphics[scale=0.75]{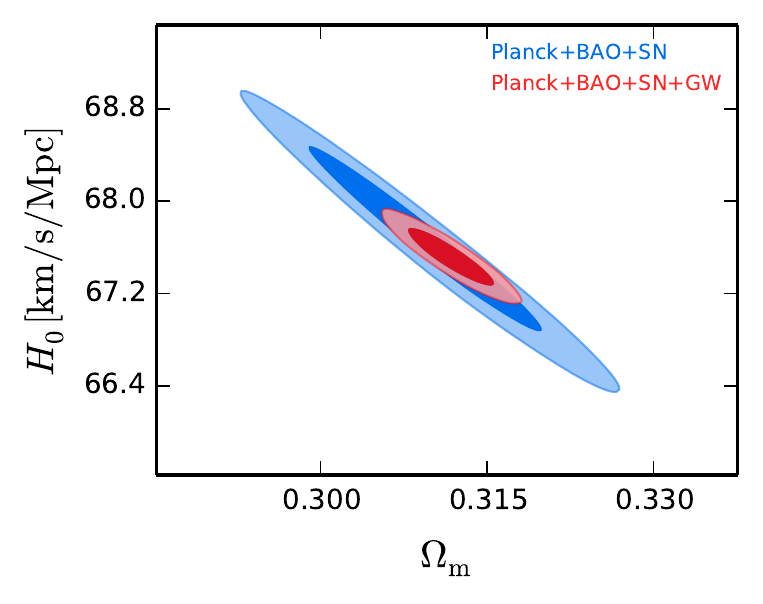}
\includegraphics[scale=0.75]{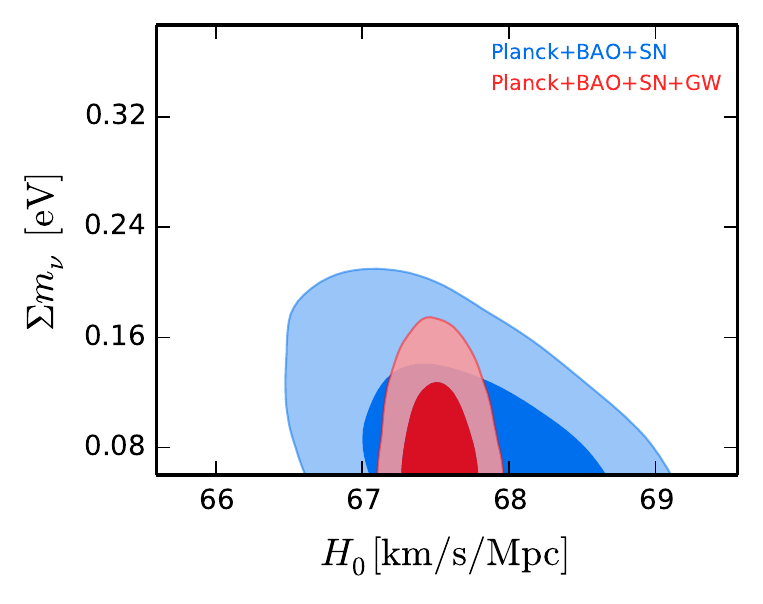}
\includegraphics[scale=0.75]{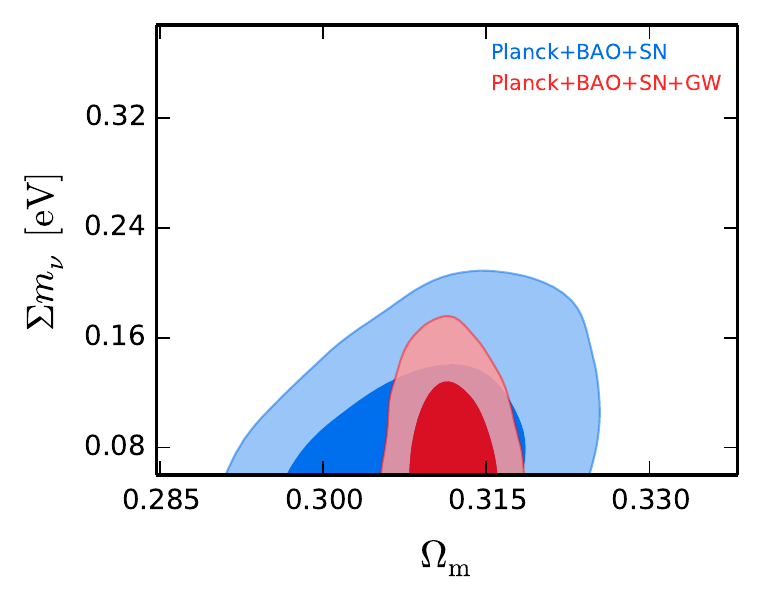}
\centering
 \caption{\label{fig2} The 68\% and 95\% CL marginalized contours of $\sum m_{\nu}$, $H_{0}$, and $\Omega_m$ using Planck+BAO+SN ({\it blue}) and Planck+BAO+SN+GW ({\it red}), for the NH case.}
\end{figure*}

\begin{figure*}[!htp]
\includegraphics[scale=0.75]{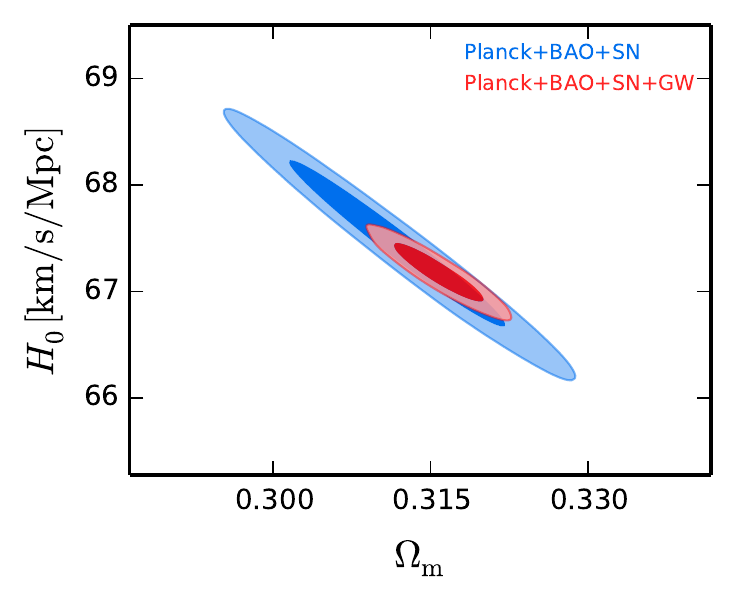}
\includegraphics[scale=0.75]{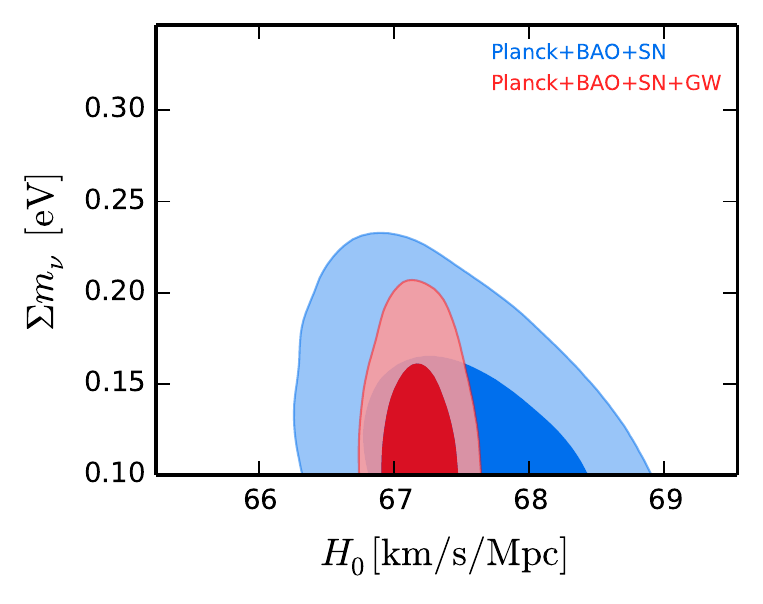}
\includegraphics[scale=0.75]{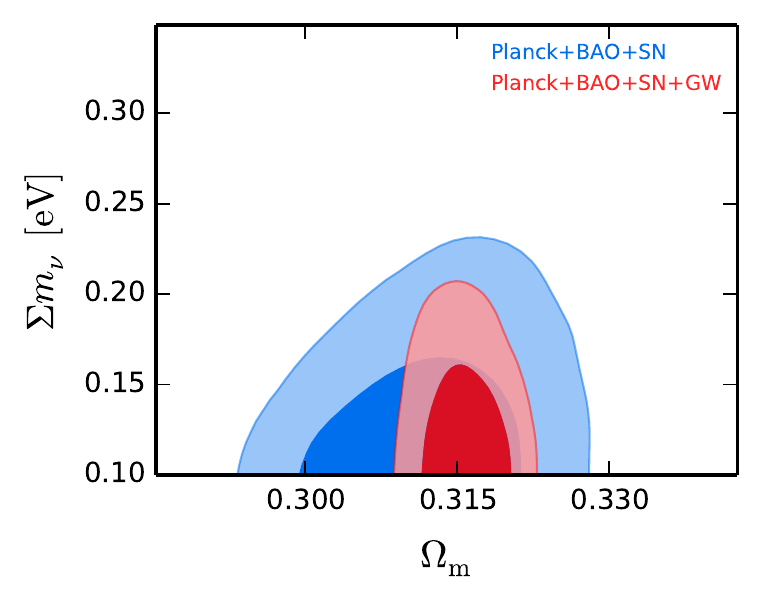}
\centering
 \caption{\label{fig3} The 68\% and 95\% CL marginalized contours of $\sum m_{\nu}$, $H_{0}$, and $\Omega_m$ using Planck+BAO+SN ({\it blue}) and Planck+BAO+SN+GW ({\it red}), for the IH case.}
\end{figure*}

\begin{figure*}[!htp]
\includegraphics[scale=0.75]{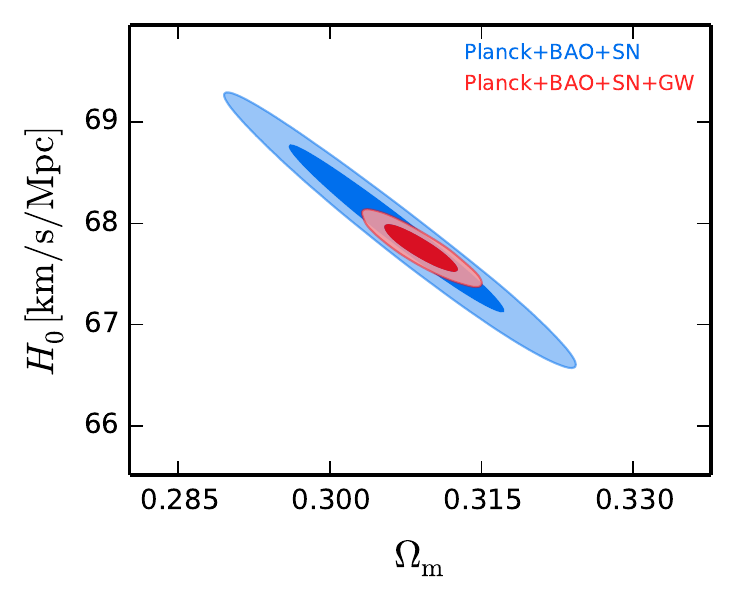}
\includegraphics[scale=0.75]{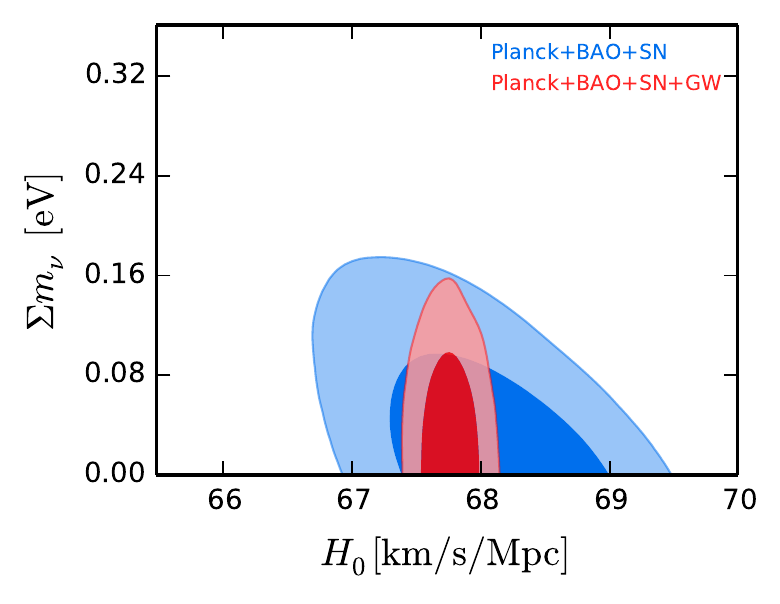}
\includegraphics[scale=0.75]{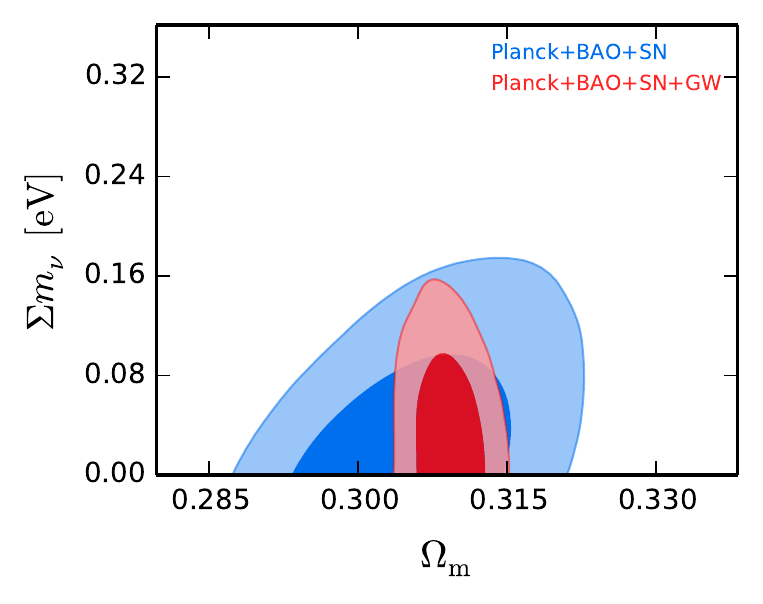}
\centering
 \caption{\label{fig4} The 68\% and 95\% CL marginalized contours of $\sum m_{\nu}$, $H_{0}$, and $\Omega_m$ using Planck+BAO+SN ({\it blue}) and Planck+BAO+SN+GW ({\it red}), for the DH case.}
\end{figure*}

Our constraint results for the neutrino mass and other cosmological parameters are shown in Table~\ref{tab1} and Figs.~\ref{fig1}--\ref{fig4}. Note that in this work we have considered three mass hierarchy cases for massive neutrinos and we have used two data combinations to make the analysis.

For convenience, we also use ``Data1" to represent the Planck+BAO+SN data combination, and use ``Data2" to represent the Planck+BAO+SN+GW data combination; e.g., see Table~\ref{tab1}. In Table~\ref{tab1}, we show the best-fit results with the $68\%$ CL uncertainties for the cosmological parameters, but owing to the fact that the neutrino mass cannot be well constrained, we only give the $95\%$ CL upper limits for the neutrino mass $\sum m_{\nu}$. In addition, the derived parameters $\Omega_m$ and $H_0$ are also listed in this table.

In Fig.~\ref{fig1}, we show the one-dimensional posterior distributions of $\sum m_{\nu}$ using the two data combinations, for the three mass hierarchy cases. In Figs.~\ref{fig2}--\ref{fig4}, we show the two-dimensional posterior distribution contours ($68\%$ and $95\%$ CL) in the $\Omega_{m}$--$H_{0}$, $H_{0}$--$\sum m_{\nu}$, and $\Omega_{m}$--$\sum m_{\nu}$ planes, for the three mass hierarchy cases, also using the two data combinations. In these figures, the blue lines and contours represent the results from the Planck+BAO+SN data, and the red lines and contours represent the results from the Planck+BAO+SN+GW data.

First, we discuss the effect of the simulated GW data of the ET on constraining the total neutrino mass. From the one-dimensional posterior distributions of $\sum m_{\nu}$ in Fig.~\ref{fig1}, we find that, when the GW data are considered, the constraints on $\sum m_{\nu}$ in all the three mass hierarchy cases become tighter, i.e., smaller values of the upper limits of $\sum m_{\nu}$ are obtained. The detailed constraint results have been given in Table~\ref{tab1}. Using the data combination Planck+BAO+SN, we obtain: $\sum m_\nu<0.175$ eV for the NH case, $\sum m_\nu<0.200$ eV for the IH case, and $\sum m_\nu<0.136$ eV for the DH case. After adding the GW data of the ET, namely when using the data combination Planck+BAO+SN+GW, the constraint results become: $\sum m_\nu<0.151$ eV for the NH case, $\sum m_\nu<0.185$ eV for the IH case, and $\sum m_\nu<0.122$ eV for the DH case. We find that the GW data help reduce the upper limits of $\sum m_\nu$ by 13.7\%, 7.5\%, and 10.3\% for the NH, IH, and DH cases, respectively. Obviously, the GW data can indeed effectively improve the constraints on the neutrino mass.

Next, we discuss how the GW data help improve the constraint accuracies for other cosmological parameters. The constraint results of the derived parameters $\Omega_m$ and $H_0$ are also listed in Table~\ref{tab1}. Using the data combination of Planck+BAO+SN, we obtain: $\Omega_m=0.3098^{+0.0064}_{-0.0065}$ and $H_{0}=67.64^{+0.50}_{-0.49}$ $\rm km~ s^{-1}~Mpc^{-1}$ for the NH case, $\Omega_m=0.3120\pm0.0064$ and $H_{0}=67.43\pm0.49$ $\rm km~ s^{-1}~Mpc^{-1}$ for the IH case, and $\Omega_m=0.3069^{+0.0065}_{-0.0071}$ and $H_{0}=67.92\pm0.52$ $\rm km~ s^{-1}~Mpc^{-1}$ for the DH case. After considering the GW data of the ET, i.e., using the data combination of Planck+BAO+SN+GW, we obtain: $\Omega_m=0.3117\pm0.0024$ and $H_{0}=67.52\pm0.16$ $\rm km~ s^{-1}~Mpc^{-1}$ for the NH case, $\Omega_m=0.3158\pm0.0026$ and $H_{0}=67.18\pm0.17$ $\rm km~ s^{-1}~Mpc^{-1}$ for the IH case, and $\Omega_m=0.3090\pm0.0023$ and $H_{0}=67.75\pm0.14$ $\rm km~ s^{-1}~Mpc^{-1}$ for the DH case. Comparing the results from the two data combinations, we find that the accuracy of $\Omega_m$ is increased by about $60\%$ and the accuracy of $H_{0}$ is increased by about $68\%$ when the GW data of the ET are considered in the cosmological fit. This indicates that the GW data of the ET can significantly improve the constraint accuracies of cosmological parameters.

We also display the two-dimensional posterior distribution contours in Figs.~\ref{fig2}--\ref{fig4}. From the blue contours (Planck+BAO+SN) in the $H_{0}$--$\sum m_{\nu}$ and $\Omega_{m}$--$\sum m_{\nu}$ planes, we can see that $\sum m_{\nu}$ is anti-correlated with $H_{0}$ and is positively correlated with $\Omega_{m}$. From the red contours (Planck+BAO+SN+GW) in the $H_{0}$--$\sum m_{\nu}$ and $\Omega_{m}$--$\sum m_{\nu}$ planes, we find that, when the GW data are considered, the parameter space is greatly shrunk in each plane and the constraints on $\Omega_{m}$ and $H_{0}$ become much tighter. Moreover, after adding the GW data, there is no obvious correlation between $\sum m_{\nu}$ and other cosmological parameter. This indicates that the degeneracies between parameters including the neutrino mass existing in the cosmological EM observations can be effectively broken by the GW observations. This is because when we use the Planck data to do the cosmological fit, the parameter combinations must be constrained to a constant $\theta_{*}$ (the observed angular size of acoustic scale $\theta_{*}=r_{s}/D_{A}$), which will cause the degeneracies between parameters. Also, neither BAO or SN can truly measure the cosmological distances ($d_A$ or $d_L$). But the GW data contain the absolute distance information at low redshifts relative to the CMB data, so they can help to break the degeneracies between $\sum m_{\nu}$ and other parameters. Hence, the upper limits on the total neutrino mass are also reduced.

To briefly summarize, our results show that the GW data can indeed improve the constraints on the total neutrino mass. When the GW data of the ET are considered, tighter bounds on the total neutrino mass could be obtained. Also, after considering the GW data, the constraints on the derived parameters $\Omega_m$ and $H_0$ become much tighter. In addition, the GW data can help break the degeneracies between $\sum m_{\nu}$ and other parameters.

\section{Conclusion}\label{sec4}

In this paper, we investigated the constraint capability of the GW observation of the ET on the total neutrino mass $\sum m_{\nu}$. We constrained the total neutrino mass in the $\Lambda$CDM cosmology by using the simulated GW data of ET in combination with other cosmological data including CMB, BAO and SN. For the three-generation neutrinos, we considered the cases of normal hierarchy, inverted hierarchy, and degenerate hierarchy. For the GW data, we considered the $\Lambda$CDM model as our fiducial model to simulate 1000 GW events that could be detected by the ET in its 10-year observation. In order to show the effect of the GW data, we used two data combinations including Planck+BAO+SN and Planck+BAO+SN+GW to constrain cosmological parameters.

Through comparing the constraint results from the two data combinations, we find that the GW data can indeed effectively improve the constraints on the total neutrino mass. Using Planck+BAO+SN, we obtain $\sum m_\nu<0.175$ eV for the NH case, $\sum m_\nu<0.200$ eV for the IH case, and $\sum m_\nu<0.136$ eV for the DH case. After considering the GW data, i.e., using Planck+BAO+SN+GW, the constraint results become $\sum m_\nu<0.151$ eV for the NH case, $\sum m_\nu<0.185$ eV for the IH case, and $\sum m_\nu<0.122$ eV for the DH case. It is found that the GW data can help reduce the upper limits of $\sum m_\nu$ by 13.7\%, 7.5\%, and 10.3\% for the NH, IH, and DH cases, respectively. For the derived parameters $\Omega_m$ and $H_0$, the GW data can significantly improve the constraint accuracies of them. The accuracy of $\Omega_m$ is increased by about $60\%$ and the accuracy of $H_{0}$ is increased by about $68\%$, when considering the GW data in the cosmological fit. In addition, the GW data can help to break the degeneracies between $\sum m_{\nu}$ and other parameters.

\begin{acknowledgments}
We would like to thank Zhou-Jian Cao, Tao Yang, Jue Zhang, and Wen Zhao for helpful discussions. This work was supported by the National Natural Science Foundation of China (Grants No.~11690021 and No.~11522540), the Top-Notch Young Talents Program of China, and the Provincial Department of Education of Liaoning (Grant No.~L2012087).

\end{acknowledgments}

%\paragraph{Note added.} This is also a good position for notes added
%after the paper has been written.


\end{document}